\documentclass[aps,twocolumn,prb,showpacs]{revtex4}
\usepackage{graphicx}
\begin{document}
\draft
\preprint{23 March 2011}
\title{Photoinduced Structural Phase Transitions in Polyacene}
\author{Shoji Yamamoto}
\address{Department of Physics, Hokkaido University,
         Sapporo 060-0810, Japan}
\date{23 March 2011}
\begin{abstract}
There exist two types of structural instability in polyacene:
double bonds in a {\it cis} pattern and those in a {\it trans} pattern.
They are isoenergetic but spectroscopically distinct.
We demonstrate optical characterization and manipulation of
Peierls-distorted polyacene employing both correlated and uncorrelated
Hamiltonians.
We clarify the phase boundaries of the {\it cis}- and
{\it trans}-distorted isomers, elucidate their optical-conductivity
spectra, and then explore their photoresponses.
There occurs a photoinduced transformation in the polyacene structure,
but it is {\it one-way} switching:
The {\it trans} configuration is well convertible into the {\it cis}
one, whereas the reverse conversion is much less feasible.
Even the weakest light irradiation can cause a transition of
uncorrelated electrons, while correlated electrons have a transition
threshold against light irradiation.
\end{abstract}
\maketitle

\section{Introduction}

   Since Salem and Longuet-Higgins pointed out that the Peierls
instability in polyacene \cite{S435} should be conditional and thus
distinct from that in polyacetylene \cite{L172}, its ground state
has been studied by numerous theoreticians with a variety of numerical
tools.
Polyacene consists of linearly fused benzene rings and is the next
simplest polymer to polyacetylene.
Nevertheless, the order of the energetic stability in its structural
isomers is still under discussion.
The {\it aromatic} (AM) configuration of regular lattice closely
competes with two types of Peierls-distorted structure: double bonds in
a {\it cis} pattern and those in a {\it trans} pattern, which read as
in-phase (IP) and out-of-phase (OP) bond order waves (BOWs) in a couple
of rows, respectively.

   Standard- and extended-H\"uckel crystalline-orbital calculations
at an early stage \cite{W23,T1069,R5430} pointed to the OP-BOW phase
as the most stable.
Band-structure calculations on a modified-neglect-of-diatomic-overlap
(MNDO) basis \cite{K305,C679} followed to find the IP-BOW phase
instead to be the most stable.
While modern computational approaches such as quantum Monte Carlo
\cite{S8927} and density-matrix renormalization group \cite{R155204}
also supported the IP-BOW phase favored over the OP-BOW one,
there exist not a few arguments \cite{B8136,G10891,S1269,L77,Y235205}
that IP and OP BOWs are almost isoenergetic.
They are indeed degenerate with each other unless we take account of
the electron itinerancy beyond the nearest-neighbor hopping
\cite{P1735} and/or the electronic correlation beyond the Hartree-Fock
(HF) scheme. \cite{Y235205}
Various {\it ab initio} calculations
\cite{C473,H5517,B7416,S045426,H134309} further show that whether and
how polyacene undergoes lattice distortion critically depend on the
supposed electronic correlation and the imposed boundary condition.
There is thus every reason to believe that the {\it cis}- and
{\it trans}-distorted isomers as well as the regular structure may
coexist due to local defects and/or thermal excitations.

   Then a query, is there any idea of designing the electronic
structure of polyacene other than energetically?
We demonstrate that {\it optics} can be the answer.
We stimulate a renewed interest in polyacene: optical observation and
manipulation of its isomers rather than energetically ranking them.

   There are transition-metal-based ladder materials and some of them
\cite{R2667,A245108,F044717} may also be interesting in the context of
varied structural instabilities.
Among others are platinum-halide double-chain compounds, where
charge-density-wave (CDW) states of the IP and OP types are both
realized.
($\mu$-bpym)[Pt(en)Cl]$_2$Cl(ClO$_4$)$_3\cdot$H$_2$O
($\mu\mbox{-bpym}=2,2'\mbox{-bipyrimidine}
 =\mbox{C}_8\mbox{H}_6\mbox{N}_4$;
 $\mbox{en}=\mbox{ethylendiamine}
 =\mbox{C}_2\mbox{H}_8\mbox{N}_2$) \cite{K7372}
and
(bpy)[Pt(dien)Br]$_2$Br$_4\cdot 2$H$_2$O
($\mbox{bpy}=4,4'\mbox{-bipyridyl}=\mbox{C}_{10}\mbox{H}_8\mbox{N}_2$;
 $\mbox{dien}=\mbox{diethylentriamine}
 =\mbox{C}_4\mbox{H}_{13}\mbox{N}_3$) \cite{K12066}
were fabricated in an attempt to bring into interaction a couple of
alternating platinum-halide linear chains via bridging organic ligands,
whose platinum ions are of mixed valence and form IP and OP CDWs,
respectively. \cite{Y235116,I063708}
The intersite electron-phonon coupling brings BOW varieties into the
hydrocarbon ladders, while the intrasite electron-phonon coupling
brings CDW varieties into the platinum-halide ladders.
Indeed the IP-CDW and OP-CDW states also exhibit distinct optical
features, \cite{Y367} but they are never yet found in a single material.
Because they are made in different energy structures, \cite{Y235116}
they are much less exchangeable with each other whether by light
irradiation \cite{Y075113} or by applying pressure \cite{Y140102}.
On the other hand, the IP-BOW and OB-BOW states in polyacene are
isoenergetic and thus able to coexist.
Here is an increasing possibility of optically tuning optical
properties.

\section{Modeling}

   In order to elucidate the electronic correlation effect on the
optical properties, we describe polyacene with and without Coulomb
interactions, generally writing the Hamiltonian as
\begin{widetext}
\begin{eqnarray}
   &&\!\!\!\!\!\!\!\!\!\!\!\!\!\!\!\!\!\!\!
   {\cal H}=
   -\sum_{l=1}^2\sum_{n=1}^N\sum_{s=\pm}
    \big[
     (t_{\parallel}-\alpha r_{l:2n-1})
     c_{l:2n-1,s}^\dagger c_{l:2n,s}
    +(t_{\parallel}-\alpha r_{l:2n})
     c_{l:2n,s}^\dagger c_{l:2n+1,s}
    +{\rm H.c.}
    \big]
   \nonumber \\
   &&\!\!\!\!\!\!\!\!\!\!\!\!\!\!\!\!\!\!\!\qquad
   -t_{\perp}\sum_{n=1}^N\sum_{s=\pm}
     \big(c_{1:2n-1,s}^\dagger c_{2:2n-1,s}+{\rm H.c.}\big)
   +\frac{K}{2}\sum_{l=1}^2\sum_{n=1}^N
     \big(r_{l:2n-1}^2+r_{l:2n}^2\big)
   +\frac{M}{2}\sum_{l=1}^2\sum_{n=1}^N
     \big(\dot{u}_{l:2n-1}^2+\dot{u}_{l:2n}^2\big)
   \nonumber \\
   &&\!\!\!\!\!\!\!\!\!\!\!\!\!\!\!\!\!\!\!\qquad
   +U\sum_{l=1}^2\sum_{n=1}^N
    \Bigl[
     \Bigl(n_{l:2n-1,+}-\frac{1}{2}\Bigr)
     \Bigl(n_{l:2n-1,-}-\frac{1}{2}\Bigr)
    +\Bigl(n_{l:2n  ,+}-\frac{1}{2}\Bigr)
     \Bigl(n_{l:2n  ,-}-\frac{1}{2}\Bigr)
    \Bigr]
   \nonumber \\
   &&\!\!\!\!\!\!\!\!\!\!\!\!\!\!\!\!\!\!\!\qquad
   +V_{\parallel}\sum_{l=1}^2\sum_{n=1}^N\sum_{s,s'=\pm}
    \Bigl(n_{l:2n,s}-\frac{1}{2}\Bigr)
    (n_{l:2n-1,s'}+n_{l:2n+1,s'}-1)
   +V_{\perp}\sum_{n=1}^N\sum_{s,s'=\pm}
    \Bigl(n_{1:2n-1,s }-\frac{1}{2}\Bigr)
    \Bigl(n_{2:2n-1,s'}-\frac{1}{2}\Bigr),
   \label{E:H}
\end{eqnarray}
\end{widetext}
where $c_{l:j,s}^\dagger$ and $c_{l:j,s}$
($c_{l:j,s}^\dagger c_{l:j,s}\equiv n_{l:j,s}$)
create and annihilate, respectively, a $\pi$ electron of spin
$s=\uparrow,\downarrow\equiv\pm$ at site $j$ on chain $l$, while
$r_{l:j}\equiv u_{l:j+1}-u_{l:j}$ signifies the bond distortion caused
by the $j$th and $(j+1)$th carbon atoms on the $l$th chain.
We assign a standard value, $2.4\,\mbox{eV}$,
\cite{S8927,R155204,R035116,S155208} for the average intrachain hopping
integral $t_{\parallel}$ and work at a sufficiently low temperature,
$k_{\rm B}T/t_{\parallel}=0.01$.
If we set the Coulomb interactions all equal to zero, eq. (\ref{E:H})
is reduced to the Su-Schrieffer-Heeger (SSH)-type Hamiltonian,
\cite{S1698} which is efficient enough to illuminate the structural
instability and excitation mechanism of polyacene.
\cite{S1269,L77,W341,A2023,A467,Z53,L477}
In the SSH modeling, the electron-lattice coupling constant $\alpha$
and the $\sigma$-bond elastic constant $K$ are usually set equal to
$4.1\,\mbox{eV}/\mbox{\AA}$ and $15.5\,\mbox{eV}/\mbox{\AA}^2$,
\cite{S1269,L77,W341} respectively, together with the carbon-site mass
$M$ of $1350\,\mbox{eV}\cdot\mbox{fs}^2/\mbox{\AA}^2$. \cite{A467,Y365}
We further set the interchain hopping integral $t_{\perp}$ equal to
$0.864t_{\parallel}$ considering that C--C bonds in the rung ($\perp$)
and leg ($\parallel$) directions are made in different lengths as
$a_{\parallel}=1.4\,\mbox{\AA}<a_{\perp}=1.5\,\mbox{\AA}$.
\cite{H134309,B7416,R035116,W5720}
The full expression (\ref{E:H}) is regarded as an extended
Peierls-Hubbard (EPH) Hamiltonian, \cite{T199} where we design the
Coulomb parameters as
$V_{\parallel(\perp)}=U/\kappa\sqrt{1+0.6117a_{\parallel(\perp)}^2}$.
\cite{O219}
$\kappa$ reads as a dielectric parameter and may range from $1$
of bare correlation \cite{R035116,S155208,S245203} to $2$ of fully
screened correlation. \cite{S155208,S245203}
Here we consider the intermediate case of $\kappa=1.5$ in an attempt to
survey potential ground states and typical photodynamics in linear
polybenzenoid hydrocarbons.

   We are interested in the generic behavior of polyacene rather than
individual features of small oligoacenes.
In view of the applicability of the periodic boundary condition and the
feasibility of simulating the electron-lattice dynamics on a large
scale, we restrict Coulomb interactions to a certain range.
We indeed learn from quantum Monte Carlo \cite{S8927} and
density-matrix renormalization-group \cite{R155204} calculations that
distant Coulomb interactions have little effect on the ground-state
properties, including the conditional Peierls instability and the
energetics of the structural isomers.
We have carried out tentative calculations of both static and dynamic
properties for $16$, $32$, $64$, and $128$ benzene rings.
Optical features such as the absolute energy and relative intensity of
every absorption are well converging even at $N=16$.
Qualitatively the same dynamics is available as a function of the
excitation density for $N\agt 32$.
Therefore, we present typical findings at $N=64$ unless otherwise noted.

\begin{figure}
\centering
\includegraphics[width=80mm]{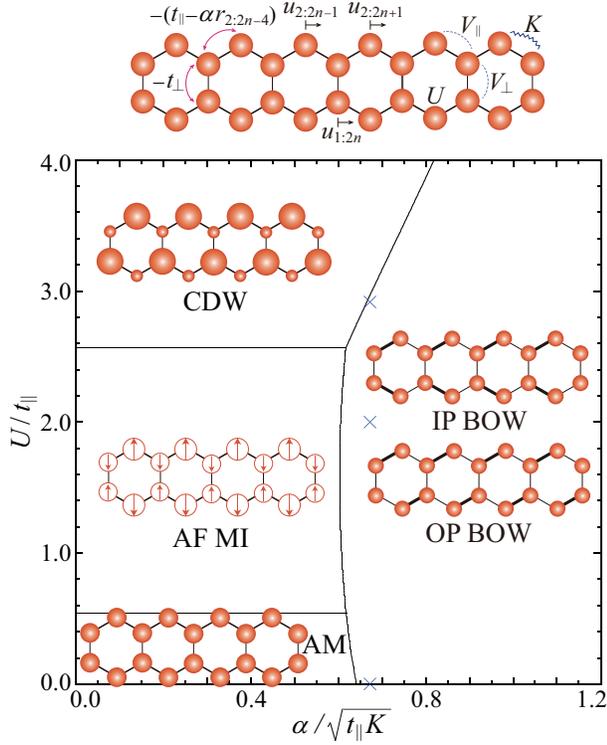}
\vspace*{-3mm}
\caption{(Color)
         A ground-state phase diagram on the $\alpha$-$U$ plain within
         the HF approximation.
         Crosses point to the SSH parameters without any correlation
         and the EPH parameters in the cases of $U=4.8\,\mbox{eV}$ and
         $U=7.0\,\mbox{eV}$.}
\label{F:PhD}
\end{figure}

\begin{figure}
\centering
\includegraphics[width=80mm]{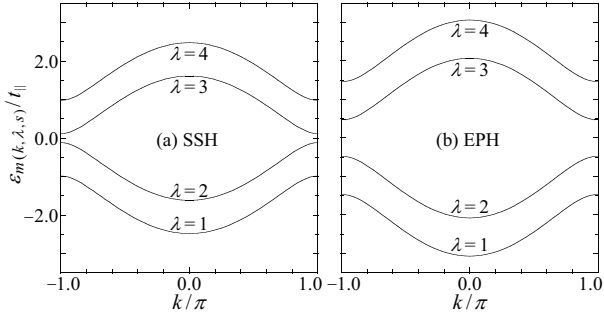}
\vspace*{-4mm}
\caption{Dispersion relations of the $\pi$-electron valence
         ($\lambda=1,2$) and conduction ($\lambda=3,4$) bands based on
         the SSH modeling (a) and the $U=4.8\,\mbox{eV}$ EPH modeling
         within the HF approximation (b).}
\vspace*{-2mm}
\label{F:disp}
\end{figure}

   Figure \ref{F:PhD} reveals possible ground states in polyacene.
When the electron-lattice and electron-electron interactions are both
weak, we find a metallic state of AM configuration.
With growing coupling and correlation, the electrons are localized into
an antiferromagnetic (AF) Mott insulator (MI) and a bond-alternating
Peierls insulator, respectively.
A CDW state also appears in the correlated region.
Such a close competition between AF MI and CDW is due to the moderately
valued screening effect.
CDW is always favored over AF MI at $\kappa=1$, \cite{K305} whereas
vice versa at $\kappa=2$. \cite{Y235205}
Though the Peierls instability in polyacene is conditional,
the well-established SSH parameters are known to result in a gapped
ground state [cf. Fig. \ref{F:disp}(a)].
The two types of Peierls insulator, IP and OP BOWs, are degenerate
in energy.
Even though Coulomb interactions are switched on, they remain
isoenergetic with their Peierls gap being enhanced
[cf. Fig. \ref{F:disp}(b)].
Keeping the SSH parameters as they are, we turn the Coulomb
interactions on.
We discuss the correlated dynamics typically at $U=4.8\,\mbox{eV}$ and
inquire further whether and how it changes getting toward a boundary to
the CDW phase.

\begin{figure}
\centering
\includegraphics[width=80mm]{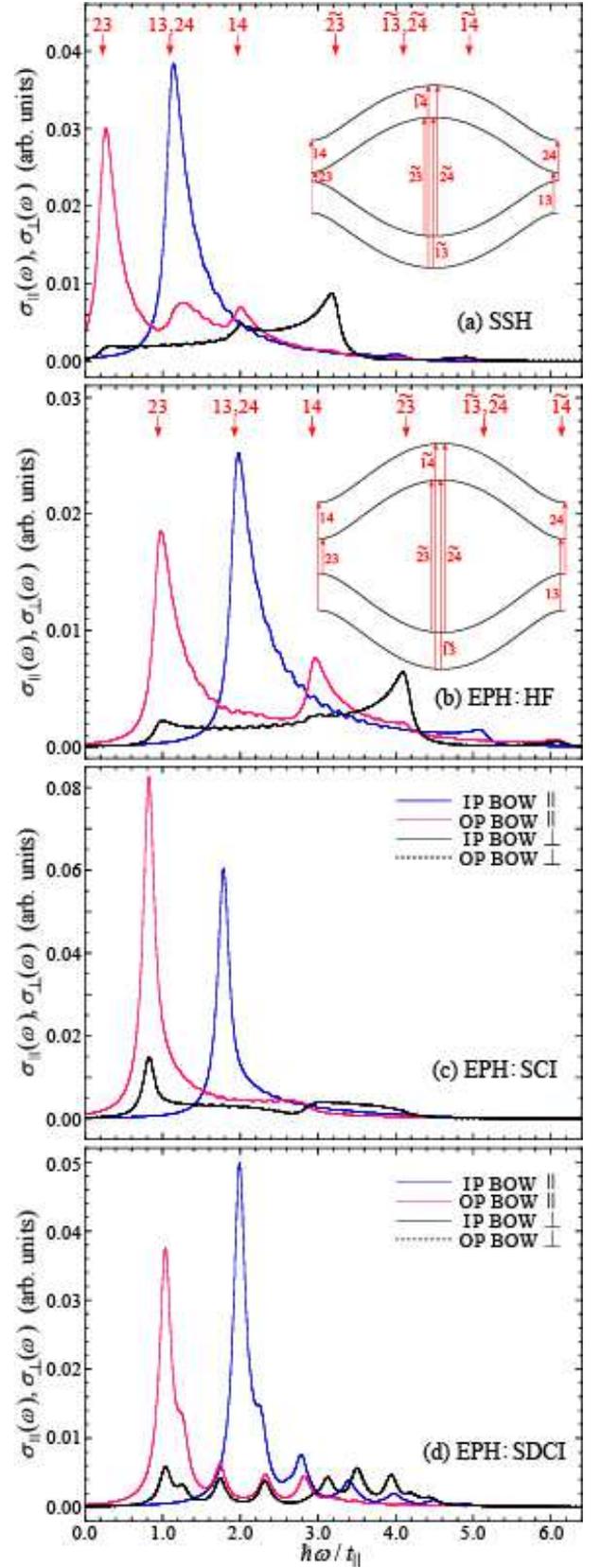}
\vspace*{-2mm}
\caption{(Color)
         Polarized optical conductivity spectra
         parallel ($\parallel$) and perpendicular ($\perp$) to the long
         axis for IP and OP BOWs, where every spectral line is
         Lorentzian broadened by a width of $0.09t_\parallel$:
         Exact calculations based on the SSH modeling (a);
         HF (b), SCI (c), and SDCI (d) calculations based on the
         $U=4.8\,\mbox{eV}$ EPH modeling.}
\label{F:OC}
\end{figure}

\section{Spectroscopic Features of the Isoenergetic Structures}

   Considering that a charge-transfer excitation costs energy of eV
order, we evaluate the real part of the optical conductivity as
\begin{equation}
   \sigma(\omega)
    =\frac{\pi}{\omega}\sum_i
   |\langle E_i|{\cal J}|E_0\rangle|^2
   \delta(E_i-E_0-\hbar\omega),
   \label{E:OC}
\end{equation}
where $|E_i\rangle$ denotes the wave vector of the $i$th-lying state
with energy $E_i$ and ${\cal J}$ signifies the current operator, which
is defined in the leg and rung directions, respectively, as
\begin{eqnarray}
   &&\!\!\!\!\!\!\!\!\!
   {\cal J}_{\parallel}=
    \frac{\sqrt{3}iea_{\parallel}}{2\hbar}\sum_{l,n,s}
    \big[
     (t_{\parallel}-\alpha r_{l:2n-1})
     c_{l:2n-1,s}^\dagger c_{l:2n,s}
   \nonumber \\
   &&\!\!\!\!\!\!\!\!\!\qquad
    +(t_{\parallel}-\alpha r_{l:2n})
     c_{l:2n,s}^\dagger c_{l:2n+1,s}
    -{\rm H.c.}
     \big],
   \\
   &&\!\!\!\!\!\!\!\!\!
   {\cal J}_{\perp}=
    \frac{iea_{\parallel}}{2\hbar}\sum_{l,n,s}(-1)^l
    \big[
     (t_{\parallel}-\alpha r_{l:2n-1})
     c_{l:2n-1,s}^\dagger c_{l:2n,s}
   \nonumber \\
   &&\!\!\!\!\!\!\!\!\!\qquad
    -(t_{\parallel}-\alpha r_{l:2n})
     c_{l:2n,s}^\dagger c_{l:2n+1,s}
    -{\rm H.c.}
     \big]
   \nonumber \\
   &&\!\!\!\!\!\!\!\!\!\qquad
   +\frac{iea_{\perp}}{\hbar}\sum_{n,s}
    t_{\perp}\big(c_{1:2n-1,s}^\dagger c_{2:2n-1,s}-{\rm H.c.}\big).
\end{eqnarray}
\begin{widetext}
\noindent
The state vector is described in terms of Slater determinants and
truncated at the order of double excitations:
\begin{eqnarray}
   &&\!\!\!\!\!\!\!
   |E_i\rangle
   =f^{(0)}(i)
    |E_0\rangle^{\quad}_{\rm HF}
   +\sum_{m(k,\mu,s)=1}^{4N}\,\sum_{m(k,\nu,s)=4N+1}^{8N}
    f^{(1)}(k,\mu,\nu,s;i)a_{m(k,\nu,s)}^\dagger a_{m(k,\mu,s)}
    |E_0\rangle^{\quad}_{\rm HF}
   \nonumber \\
   &&\!\!\!\!\!\!\!\qquad
   +\sum_{m(k_1,\mu_1,s_1)>m(k_2,\mu_2,s_2)=1}^{4N}\,
    \sum_{m(k_1,\nu_1,s_1)>m(k_2,\nu_2,s_2)=4N+1}^{8N}
    f^{(2)}(k_1,k_2,\mu_1,\mu_2,\nu_1,\nu_2,s_1,s_2;i)
   \nonumber \\
   &&\!\!\!\!\!\!\!\qquad\times
    a_{m(k_1,\nu_1,s_1)}^\dagger a_{m(k_2,\nu_2,s_2)}^\dagger
    a_{m(k_1,\mu_1,s_1)} a_{m(k_2,\mu_2,s_2)}
    |E_0\rangle^{\quad}_{\rm HF},
\end{eqnarray}
\end{widetext}
where
$|E_0\rangle^{\quad}_{\rm HF}
 \equiv\prod_{m=1}^{4N} a_{m}^\dagger|0\rangle$
is the ground-state HF wave function with $|0\rangle$ being the true
electron vacuum and $a_m^\dagger$ creating an electron in the $m$th HF
orbital with energy $\varepsilon_m$.
The orbital counter $m(k,\lambda,s)$ is composed of
the wave number ($k=2\pi n/N;\,n=1,\cdots,N$),
band label ($\lambda=1,\cdots,4$) and
spin quantum number ($s=\pm$).
Since an electron excitation is optically allowed without any momentum
transfer, \cite{H1788} the periodic boundary condition serves to reduce
the number of excited states to be considered.

   Excited states of the HF type consist of a single Slater
determinant, \cite{H1376,O045122} where
$f^{(0)}(i)=\delta_{0,i}$,
$f^{(1)}(k,\mu,\nu,s;i)=\delta_{k\mu\nu s,i}$ and
$f^{(2)}(k_1,k_2,\mu_1,\mu_2,\nu_1,\nu_2,s_1,s_2;i)=0$, giving
$E_i=\!\!\!\!\!\!\quad^{\quad}_{\rm HF\!}\langle E_0|
     {\cal H}|E_0\rangle^{\quad}_{\rm HF}
    -\varepsilon_{m(k,\mu,s)}+\varepsilon_{m(k,\nu,s)}$.
Those of the CI type read as resonating Slater determinants,
\cite{S125204} where their coefficients are determined so as to
diagonalize the original Hamiltonian (\ref{E:H}).
Within the single-excitation CI (SCI) scheme,
$f^{(0)}(i)$ and $f^{(2)}(k_1,k_2,\mu_1,\mu_2,\nu_1,\nu_2,s_1,s_2;i)$
remain unchanged and therefore only excited states are improved.
Here is no correction to the ground state, because no single-excitation
Slater determinant is obtained by the Hamiltonian operation on the HF
ground state.
Only configurations of the double-excitation type can {\it directly}
interact with the HF ground state.
Therefore, it pays most to truncate the series expansion at the present
order, which is referred to as the single-double-excitation CI (SDCI)
scheme, though the cost of SDCI ($\sim N^6$) is incomparably larger
than those of HF ($\sim N^3$) and SCI ($\sim N^4$).

   Figure \ref{F:OC} shows comparative calculations of the optical
conductivity spectra for IP and OP BOWs, where only the SDCI
calculation is carried out at $N=16$.
For the light polarized perpendicular to the conjugation direction,
the two isoenergetic structures still look alike even in the SDCI
scheme.
The valence band of $\lambda=1$ and the conduction band of $\lambda=3$
are both composed of molecular orbitals symmetric with respect to the
reflection about the plane bisecting every rung bond, whereas those of
$\lambda=2$ and $\lambda=4$ consist of molecular orbitals antisymmetric
to the reflection.
Such properties of PM remain unchanged with bond distortion, whether
the {\it cis} or {\it trans} type, while they do not hold in AF MI
and CDW.
Therefore, interband electronic excitations of the $1$-to-$3$ and
$2$-to-$4$ types make no contribution to the rung-direction optical
conductivity in both IP and OP BOWs.
On the other hand, a long-axis-polarized photon clearly distinguishes
between IP and OP BOWs.
In the conjugation direction, $2$-to-$3$ interband excitations,
including the highest-occupied-molecular-orbital
(HOMO)-to-lowest-unoccupied-molecular-orbital (LUMO) transition, are
optically forbidden on an IP-BOW background but allowed on an OP-BOW
one.
OP BOW can thus absorb long-axis-polarized photons of much lower energy
than IP BOW can.
The zone-center ($k=\pi$) $1$-to-$3$ and $2$-to-$4$ interband
excitations are most contributive to the IP-BOW spectra in the
conjugation direction, whether correlated or uncorrelated, while the
OP-BOW spectra in the conjugation direction qualitatively vary with the
relevant electronic correlation.
Valence-to-conduction-band excitations of any type are optically
allowed at the uncorrelated level, but only near HOMO-to-LUMO
transitions survive the strong correlation.

   IP and OP BOWs may be coexistent in actual oligoacenes.
They both exhibit selective absorption of their own for the light
polarized parallel to the conjugation direction, especially under the
influence of the electronic correlation.
Then it is highly interesting whether and how the isoenergetic
structural isomers are photoactive.
Sufficiently oriented long acenes may have an application to the
optical memory.

\section{Photoinduced Phase Transitions between the Isoenergetic
         Structures}

   We simulate time evolution of a charge-transfer excitation solving
the Schr\"odinger equation
\begin{equation}
   i\hbar{\dot{\mit\Psi}}_s(t)
  ={\cal H}_s(t){\mit\Psi}_s(t),
   \label{E:Schrodinger}
\end{equation}
where the spin-$s$ sector of the Hamiltonian, ${\cal H}_s(t)$, and a
complete set of its wave functions, ${\mit\Psi}_s(t)$, are given by
square matrices of degree $4N$.
The SSH Hamiltonian is exactly computable, while the EPH Hamiltonian is
treated within the HF approximation.
In both cases, defining the time-dependent wave functions as
\begin{equation}
   {\mit\Psi}_s(t)
  =\left[
    \begin{array}{ccc}
     \psi_{1: 1, 1,s}(t) & \cdots & \psi_{1: 1,4N,s}(t) \\
     \psi_{2: 1, 1,s}(t) & \cdots & \psi_{2: 1,4N,s}(t) \\
     \vdots & & \vdots \\
     \psi_{1:2N, 1,s}(t) & \cdots & \psi_{1:2N,4N,s}(t) \\
     \psi_{2:2N, 1,s}(t) & \cdots & \psi_{2:2N,4N,s}(t)
    \end{array}
   \right],
   \label{E:Psi}
\vspace*{-1mm}
\end{equation}
eq. (\ref{E:Schrodinger}) is explicitly written as
\begin{widetext}
\begin{eqnarray}
   &&\!\!\!\!\!\!\!\!\!\!\!\!\!\!\!
   i\hbar\dot{\psi}_{1:2n-1,\iota,s}(t)
  =-\big[t_\perp+V_\perp p_{2n-1,s}^{\perp\,*}(t)\big]
    \psi_{2:2n-1,\iota,s}(t)
   \nonumber \\
   &&\!\!\!\!\!\!\!\!\!\!\!\!\!\!\!\qquad
   -\big[
     t_\parallel-\alpha r_{1:2n-2}(t)
    +V_\parallel p_{1:2n-2,s}^{\parallel}(t)
    \big]\psi_{1:2n-2,\iota,s}(t)
   -\big[
     t_\parallel-\alpha r_{1:2n-1}(t)
    +V_\parallel p_{1:2n-1,s}^{\parallel\,*}(t)
    \big]\psi_{1:2n,\iota,s}(t)
   \nonumber \\
   &&\!\!\!\!\!\!\!\!\!\!\!\!\!\!\!\qquad
   +\Bigl\{
    U\big[d_{1:2n-1,-s}(t)-\frac{1}{2}\big]
   +V_\parallel\big[d_{1:2n-2}(t)+d_{1:2n}(t)-2\big]
   +V_\perp\big[d_{2:2n-1}(t)-1\big]
    \Bigr\}\psi_{1:2n-1,\iota,s}(t),
   \nonumber \\
   &&\!\!\!\!\!\!\!\!\!\!\!\!\!\!\!
   i\hbar\dot{\psi}_{2:2n-1,\iota,s}(t)
  =-\big[t_\perp+V_\perp p_{2n-1,s}^{\perp}(t)\big]
    \psi_{1:2n-1,\iota,s}(t)
   \nonumber \\
   &&\!\!\!\!\!\!\!\!\!\!\!\!\!\!\!\qquad
   -\big[
     t_\parallel-\alpha r_{2:2n-2}(t)
    +V_\parallel p_{2:2n-2,s}^{\parallel}(t)
    \big]\psi_{1:2n-2,\iota,s}(t)
   -\big[
     t_\parallel-\alpha r_{2:2n-1}(t)
    +V_\parallel p_{2:2n-1,s}^{\parallel\,*}(t)
    \big]\psi_{2:2n,\iota,s}(t)
   \nonumber \\
   &&\!\!\!\!\!\!\!\!\!\!\!\!\!\!\!\qquad
   +\Bigl\{
    U\big[d_{2:2n-1,-s}(t)-\frac{1}{2}\big]
   +V_\parallel\big[d_{2:2n-2}(t)+d_{2:2n}(t)-2\big]
   +V_\perp\big[d_{1:2n-1}(t)-1\big]
    \Bigr\}\psi_{2:2n-1,\iota,s}(t),
   \nonumber \\
   &&\!\!\!\!\!\!\!\!\!\!\!\!\!\!\!
   i\hbar\dot{\psi}_{1:2n,\iota,s}(t)
  =-\big[
     t_\parallel-\alpha r_{1:2n-1}(t)
    +V_\parallel p_{1:2n-1,s}^{\parallel}(t)
    \big]\psi_{1:2n-1,\iota,s}(t)
   -\big[
     t_\parallel-\alpha r_{1:2n}(t)
    +V_\parallel p_{1:2n,s}^{\parallel\,*}(t)
    \big]\psi_{1:2n+1,\iota,s}(t)
   \nonumber \\
   &&\!\!\!\!\!\!\!\!\!\!\!\!\!\!\!\qquad
   +\Bigl\{
    U\big[d_{1:2n,-s}(t)-\frac{1}{2}\big]
   +V_\parallel\big[d_{1:2n-1}(t)+d_{1:2n+1}(t)-2\big]
    \Bigr\}\psi_{1:2n,\iota,s}(t),
   \nonumber \\
   &&\!\!\!\!\!\!\!\!\!\!\!\!\!\!\!
   i\hbar\dot{\psi}_{2:2n,\iota,s}(t)
  =-\big[
     t_\parallel-\alpha r_{2:2n-1}(t)
    +V_\parallel p_{2:2n-1,s}^{\parallel}(t)
    \big]\psi_{2:2n-1,\iota,s}(t)
   -\big[
     t_\parallel-\alpha r_{2:2n}(t)
    +V_\parallel p_{2:2n,s}^{\parallel\,*}(t)
    \big]\psi_{2:2n+1,\iota,s}(t)
   \nonumber \\
   &&\!\!\!\!\!\!\!\!\!\!\!\!\!\!\!\qquad
   +\Bigl\{
    U\big[d_{2:2n,-s}(t)-\frac{1}{2}\big]
   +V_\parallel\big[d_{2:2n-1}(t)+d_{2:2n+1}(t)-2\big]
    \Bigr\}\psi_{2:2n,\iota,s}(t),
\end{eqnarray}
\end{widetext}
where the electron-density and bond-order parameters are calculated
through
\begin{eqnarray}
   &&\!\!\!\!\!\!\!\!\!\!\!\!\!\!\!\!\!\!\!
   d_{l:j}(t)
  =\sum_s d_{l:j,s}(t)
  =\sum_{s}\mathop{{\sum}'}_{\!\iota}
   \big|\psi_{l:j,\iota,s}(t)\big|^2,
   \label{E:d}
   \\
   &&\!\!\!\!\!\!\!\!\!\!\!\!\!\!\!\!\!\!\!
   p_{l:j}^\parallel(t)
  =\sum_s p_{l:j,s}^\parallel(t)
  =\sum_{s}\mathop{{\sum}'}_{\!\iota}
   \psi_{l:j,\iota,s}^*(t)\psi_{l:j+1,\iota,s}(t),
   \label{E:pleg}
   \\
   &&\!\!\!\!\!\!\!\!\!\!\!\!\!\!\!\!\!\!\!
   p_{j}^\perp(t)
  =\sum_s p_{j,s}^\perp(t)
  =\sum_{s}\mathop{{\sum}'}_{\!\iota}
   \psi_{1:j,\iota,s}^*(t)\psi_{2:j,\iota,s}(t),
   \label{E:prung}
\end{eqnarray}
with $\sum^\prime$ denoting a summation over the occupied levels
immediately after the photoexcitation.

   Discretizing the time variable as
$t_m=m{\mit\Delta}t$ $(m=0,1,2,\cdots)$
with an interval much smaller than the optical-phonon time scale, say,
${\mit\Delta}t=10^{-3}\sqrt{M/K}\simeq 0.01\,\mbox{fs}$, we integrate
eq. (\ref{E:Schrodinger}) step by step:
\begin{equation}
   {\mit\Psi}_s(t_{m+1})
  ={\rm exp}
   \Bigl[
   -\frac{i{\cal H}_s(t_m)}{\hbar}{\mit\Delta}t
   \Bigr]
   {\mit\Psi}_s(t_m).
   \label{E:integration}
\end{equation}
At $t=t_0$, the system is {\it photoexcited},
that is, some valence electrons are pumped up into a conduction band
across the Peierls gap.
In response to the photoinduced interband electronic transition, we
{\it immediately} construct the initial Hamiltonian ${\cal H}_s(t_0)$
from the modified electronic parameters
$d_{l:j}(t_0)$, $p_{l:j}^\parallel(t_0)$, and $p_{j}^\perp(t_0)$
leaving the background lattice untouched.
Then ${\mit\Psi}_s(t_0)^\dagger{\cal H}_s(t_0){\mit\Psi}_s(t_0)$ is
no more diagonal and that is why the wave vectors begin to fluctuate.
Obtaining the instantaneous eigenfunctions and eigenvalues
\begin{eqnarray}
   &&\!\!\!\!\!\!\!\!\!\!\!\!\!\!\!\!\!\!\!
   {\mit\Phi}_s(t_m)
  =\left[
    \begin{array}{ccc}
     \phi_{1: 1, 1,s}(t_m) & \cdots & \phi_{1: 1,4N,s}(t_m) \\
     \phi_{2: 1, 1,s}(t_m) & \cdots & \phi_{2: 1,4N,s}(t_m) \\
     \vdots & & \vdots \\
     \phi_{1:2N, 1,s}(t_m) & \cdots & \phi_{1:2N,4N,s}(t_m) \\
     \phi_{2:2N, 1,s}(t_m) & \cdots & \phi_{2:2N,4N,s}(t_m)
    \end{array}
   \right],
   \label{E:Psi}
   \\
   &&\!\!\!\!\!\!\!\!\!\!\!\!\!\!\!\!\!\!\!
   {\mit\Phi}_s^\dagger(t_m){\cal H}_s(t_m){\mit\Phi}_s(t_m)
  =\left[
    \begin{array}{ccc}
     \varepsilon_{    1,s}(t_m) & & \\
     & \ddots &                     \\
     & & \varepsilon_{   4N,s}(t_m) \\
    \end{array}
   \right],
   \label{E:varepsilon}
\end{eqnarray}
from  ${\cal H}_s(t_m)$ at every time step, \cite{Y415215} we can
express eq. (\ref{E:integration}) as
\begin{eqnarray}
   &&\!\!\!\!\!\!\!\!\!
   \psi_{l':j',\iota',s}(t_{m+1})
  =\sum_{\iota=1}^{4N}\sum_{l=1}^2\sum_{j=1}^{2N}
   {\rm exp}
   \Bigl[
   -\frac{i\varepsilon_{\iota,s}(t_m)}{\hbar}{\mit\Delta}t
   \Bigr]
   \nonumber \\
   &&\!\!\!\!\!\!\!\!\!\qquad\times
   \phi_{l':j',\iota,s}(t_m)
   \phi_{l:j,\iota,s}^*(t_m)
   \psi_{l:j,\iota',s}(t_m).
\end{eqnarray}

   Once the electronic wave functions deviate from the ground-state
equilibrium, the lattice begins to fluctuate correspondingly.
Its dynamics is governed by Newton's equation of motion,
\begin{eqnarray}
   &&\!\!\!\!\!\!\!\!\!
   M{\ddot u}_{l:j}(t)
  =K\big[u_{l:j-1}(t)-2u_{l:j}(t)+u_{l:j-1}(t)\big]
   \nonumber \\
   &&\!\!\!\!\!\!\!\!\!\qquad
  +2\alpha{\rm Re}\big[p_{l:j}^\parallel(t)-p_{l:j-1}^\parallel(t)\big]
   \equiv F_{l:j}(t),
   \label{E:Newton}
\end{eqnarray}
and solvable stepwise:
\begin{eqnarray}
   &&\!\!\!\!\!\!\!\!\!
   \dot{u}_{l:j}(t_{m+1})
  =\dot{u}_{l:j}(t_{m})+\frac{F_{l:j}(t_{m})}{M}{\mit\Delta}t,
   \nonumber \\
   &&\!\!\!\!\!\!\!\!\!
   u_{l:j}(t_{m+1})
  =u_{l:j}(t_{m})+\dot{u}_{l:j}(t_{m}){\mit\Delta}t.
\end{eqnarray}
The classical treatment of lattice degrees of freedom is successful in
investigating both correlated \cite{M2282} and uncorrelated
\cite{I024302} electron-lattice dynamics.
We avoid setting any ``artificial" fluctuation on the lattice initially
but start every calculation from a stationary BOW lattice of the IP
($\sigma=1$) or OP ($\sigma=-1$) type:
\begin{equation}
   \sigma^l r_{l:j}(t_0)=-\sigma^l r_{l:j+1}(t_0),\ \ 
   \dot{u}_{l:j}(t_0)=0,
   \label{E:initial}
\end{equation}
in an attempt to simulate every precursor to structural phase
transitions quantitatively.
\begin{figure*}
\centering
\includegraphics[width=170mm]{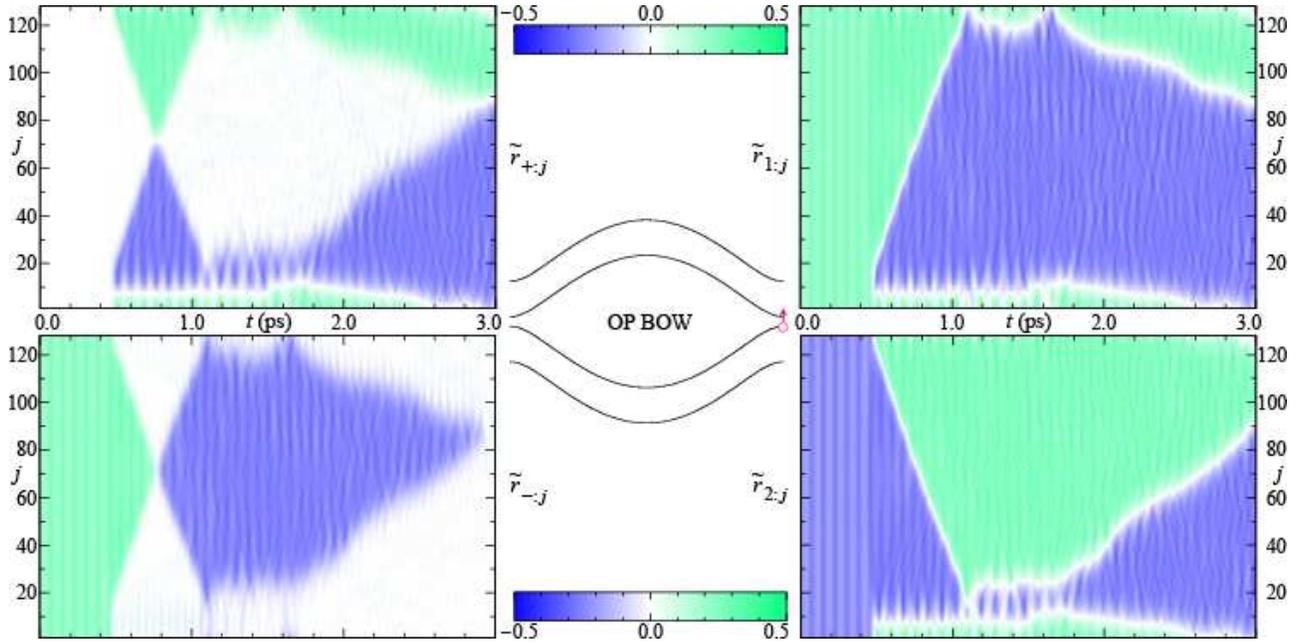}
\vspace*{-2mm}
\caption{(Color)
         SSH-model calculations of the bond variables
         $\widetilde{r}_{\pm:j}$ and $\widetilde{r}_{l:j}$
         as functions of space $j$ and time $t$ in the case of
         the HOMO-to-LUMO one-electron excitation on the OP-BOW
         background.}
\label{F:SSHdynOP23-1}
\end{figure*}
\begin{figure*}
\centering
\includegraphics[width=170mm]{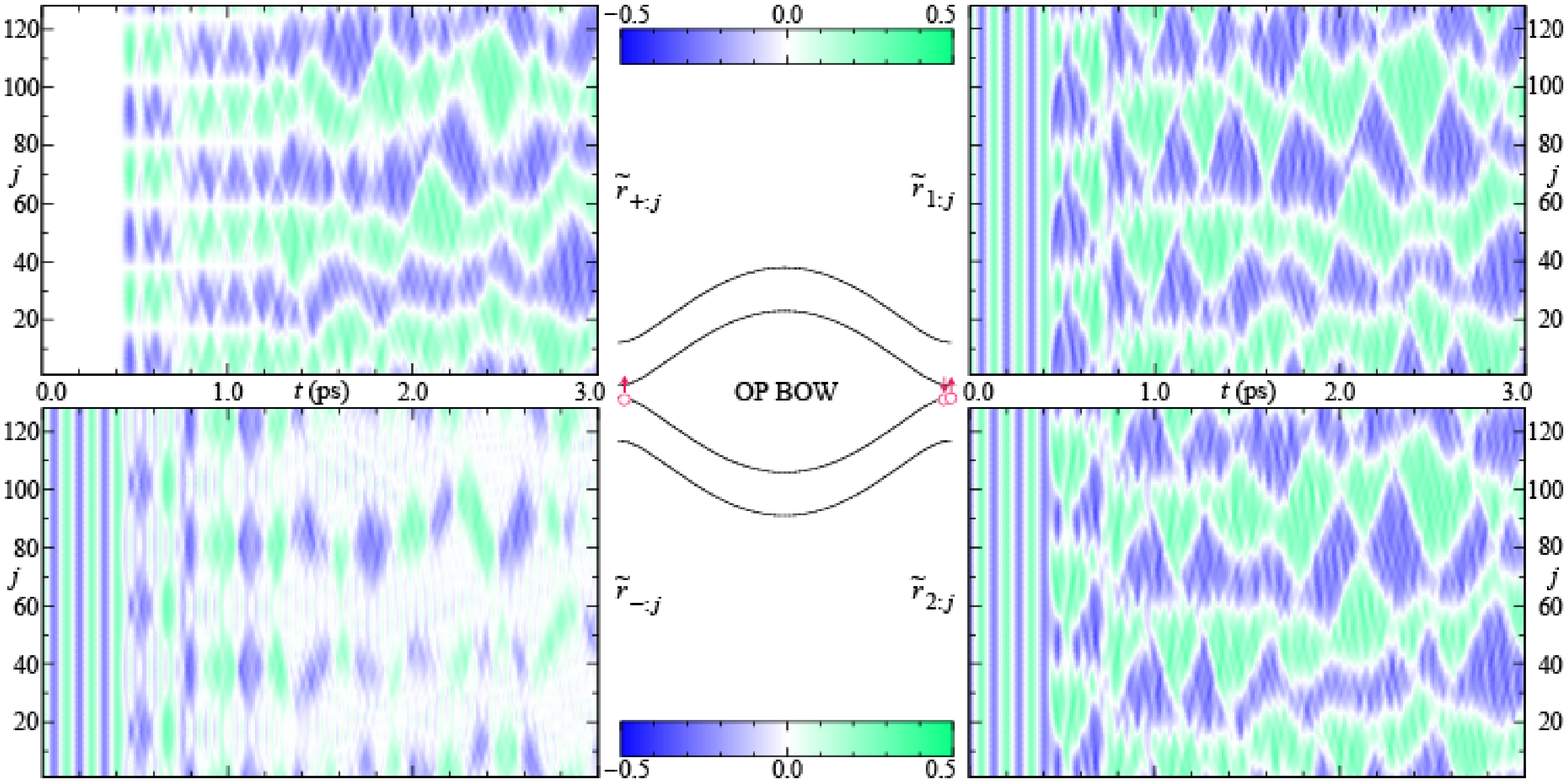}
\vspace*{-2mm}
\caption{(Color)
         The same as Fig. \ref{F:SSHdynOP23-1} but in the case of a
         $2$-to-$3$ interband three-electron excitation at
         $k\simeq\pi$.}
\label{F:SSHdynOP23-3}
\end{figure*}

   In order to visualize bond arrangements of the IP and OP types, we
introduce effective bond variables,
$(\widetilde{r}_{1:j}\pm\widetilde{r}_{2:j})/2
 \equiv\widetilde{r}_{\pm:j}$,
where the bare bond distortions $r_{l:j}$ are decomposed into their
net ($\bar{r}_{l:j}$) and alternating ($\widetilde{r}_{l:j}$)
components as
$\bar{r}_{l:j}=(2r_{l:j}+r_{l:j-1}+r_{l:j+1})/4$ and
$\widetilde{r}_{l:j}=(2r_{l:j}-r_{l:j-1}-r_{l:j+1})/4$.
$\widetilde{r}_{l:j}$ remaining constant with varying $j$ read as
regular bond alternation on the $l$th chain, while a change of sign in
them denotes a kink (domain wall) occurring.
$\widetilde{r}_{+:j}$ and $\widetilde{r}_{-:j}$ serve as detectors of
the IP-BOW and OP-BOW configurations, respectively.
Now we are ready to observe photoexcitations of polyacene.
We consider, unless otherwise noted, a few electrons being excited by
around Peierls-gap energy, that is, transitions from the highest one or
more occupied to the lowest one or more unoccupied molecular orbitals.
The thus-induced electron-lattice dynamics depends on the momentum and
spin of the excited electrons but exhibits universal features.
\begin{figure*}
\centering
\includegraphics[width=170mm]{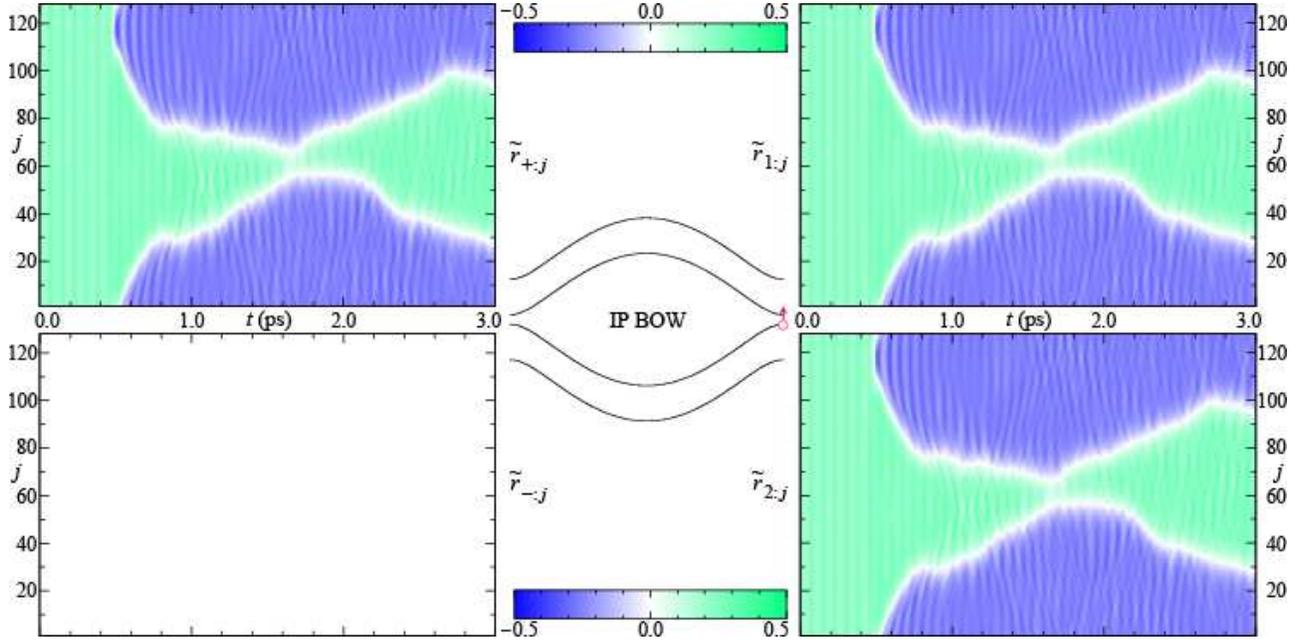}
\vspace*{-2mm}
\caption{(Color)
         The same as Fig. \ref{F:SSHdynOP23-1} but on the IP-BOW
         background.}
\label{F:SSHdynIP23-1}
\end{figure*}
\begin{figure*}
\centering
\includegraphics[width=170mm]{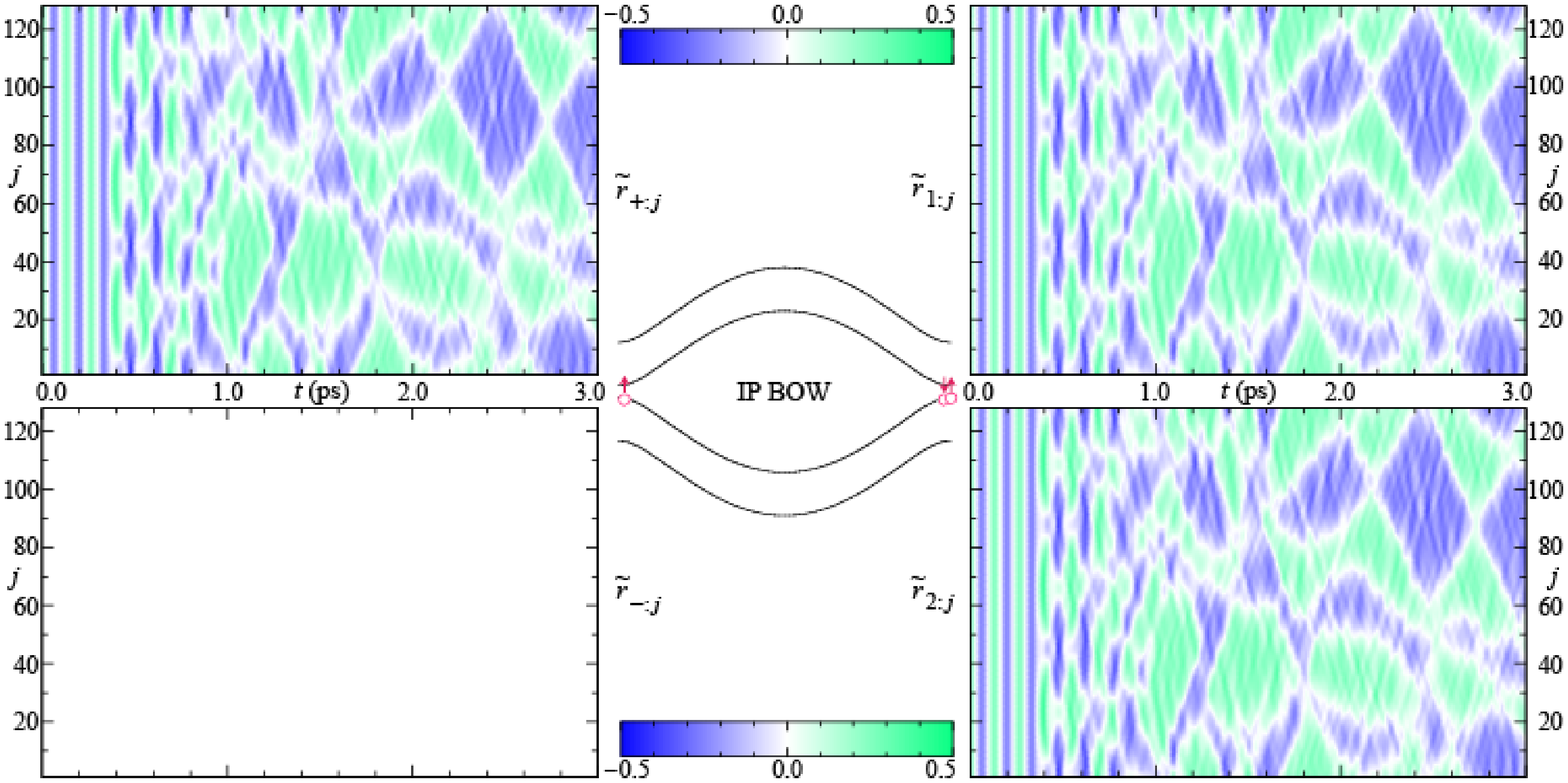}
\vspace*{-2mm}
\caption{(Color)
         The same as Fig. \ref{F:SSHdynOP23-3} but on the IP-BOW
         background.}
\label{F:SSHdynIP23-3}
\end{figure*}

\begin{figure*}
\centering
\includegraphics[width=170mm]{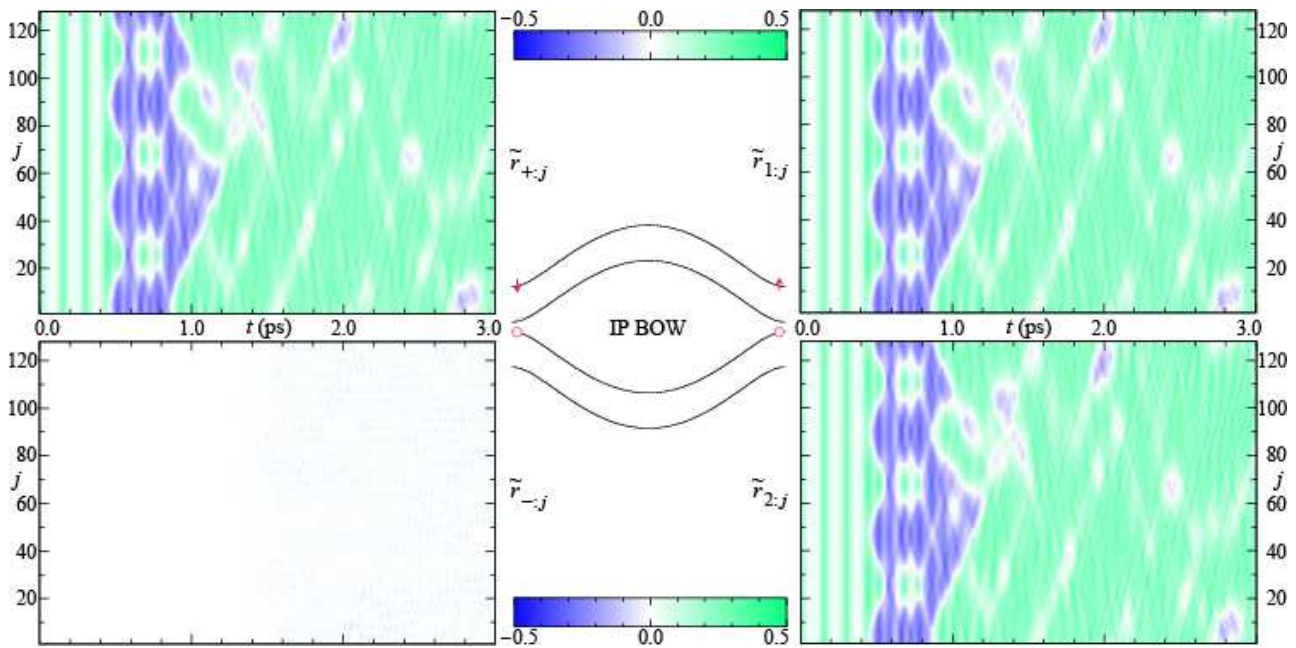}
\vspace*{-2mm}
\caption{(Color)
         The same as Fig. \ref{F:SSHdynIP23-3} but in the case of a
         $2$-to-$4$ interband two-electron excitation at $k\simeq\pi$.}
\label{F:SSHdynIP24-2}
\end{figure*}

\subsection{Uncorrelated Electron-Lattice Dynamics}

   We begin with correlation-free photoresponses.
In Fig. \ref{F:SSHdynOP23-1} we demonstrate the most fundamental
HOMO-to-LUMO one-electron excitation of the OP-BOW ground state, which
is optically allowed in both leg and rung directions.
The photon energy is first spent on a background uniform oscillation
and then contributes to an OP-to-IP-BOW phase transition.
The two bond-ordered chains are disarranged at the same time.
The defects in both chains split into a pair of kink and antikink.
Then the resultant antidomains grow ``repulsively".
The {\it out-of-phase} lattice fluctuations convert the OP-BOW
arrangement into the IP-BOW one.
The OP-BOW configuration macroscopically recovers within a picosecond
but declines again, and finally disappears into the global IP-BOW
configuration.
Stronger light irradiation drives more electrons to jump over the
Peierls gap, induces a precursory global oscillation of larger
amplitude and longer period, and then generates multipairs of kink and
antikink against the initial bond arrangement.
Figure \ref{F:SSHdynOP23-3} shows the lattice relaxation from a
low-lying three-electron excited state.
Declining and recovering again and again in such a complicated manner,
OP BOW disappears into IP BOW within a few picoseconds.
Such a swinging motion is observed in a double-chain CDW system
without any Coulomb interaction as well \cite{I060302} and therefore
recognized to be characteristic of uncorrelated electrons.

   What happens to IP BOW being photoexcited in the same way?
Figure \ref{F:SSHdynIP23-1} demonstrates the HOMO-to-LUMO one-electron
excitation of the IP-BOW ground state and the following lattice
relaxation.
Indeed the absorbed photon similarly brings a pair of kink and antikink
to both chains, but the resultant antidomains now grow ``attractively".
The kink-antikink breathing motions in the two chains are completely
{\it in phase} and there is no sign of OP BOW appearing.
IP BOW withstands stronger light irradiation.
Figure \ref{F:SSHdynIP23-3} shows that IP BOW with three electrons
being excited still keeps itself within a configuration of the IP type.
Pumping further electrons up to the lower conduction band, we find no
sign of OP-BOW nucleation, even though numbers of antidomains eat away
into the initial configuration.
Thus excited IP BOW is expected to relax, whether radiatively or
nonradiatively, into the initial state via kink-antikink geminate
recombinations.

   We have so far investigated low-energy excitations of near Peierls
gap.
OP BOW can be converted to IP BOW by the weakest light irradiation,
while IP BOW is light-resistant and survives much stronger
photoirradiation.
However, there is a possibility of such an inactive response of the
IP-BOW state resulting from the light polarization.
We should be reminded that a long-axis-polarized photon is not
contributive to any interband electronic excitation of the $2$-to-$3$
type on an IP-BOW background (cf. Fig. \ref{F:OC}).
Therefore, Figs. \ref{F:SSHdynIP23-1} and \ref{F:SSHdynIP23-3} just say
that charge-transfer excitations in the rung direction hardly lead to
an IP-to-OP bond-structure conversion.
We should inquire further into a possibility of phototuning
IP BOW by higher-energy excitations.
We thus try the $2$-to-$4$ interband one-electron excitation at $k=\pi$,
which reads as the optically-allowed lowest-energy charge-transfer
excitation in the conjugation direction on an IP-BOW background, but
the resultant lattice motion is almost the same as
Fig. \ref{F:SSHdynIP23-1} without any sign of OP-BOW configuration
appearing.
When we excite further electrons to the upper conduction band
(cf. Fig. \ref{F:SSHdynIP24-2}), two chains are no more exactly
harmonized with each other.
Transient islands of OP-BOW configuration come to appear here and there
with increasing number of electrons being excited, but they do not grow
up completely.
IP BOW looks so persistent and tough against any light irradiation.
\begin{figure*}
\centering
\includegraphics[width=170mm]{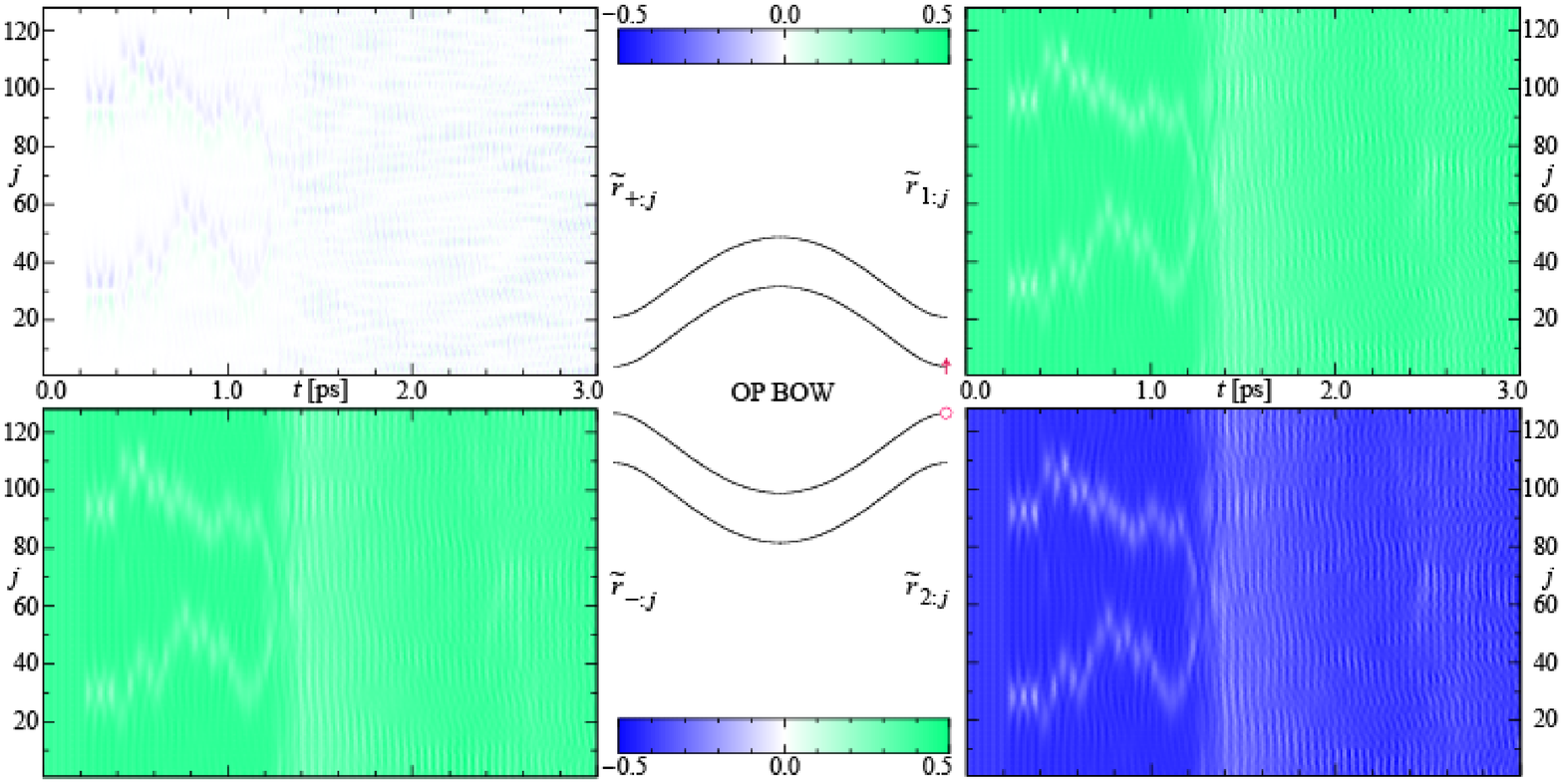}
\vspace*{-2mm}
\caption{(Color)
         $U=4.8\,\mbox{eV}$-EPH-model calculations of the bond variables
         $\widetilde{r}_{\pm:j}$ and $\widetilde{r}_{l:j}$
         as functions of space $j$ and time $t$ in the case of
         the HOMO-to-LUMO one-electron excitation on the OP-BOW
         background.}
\label{F:EPH4.8eVdynOP23-1}
\end{figure*}
\begin{figure*}
\centering
\includegraphics[width=170mm]{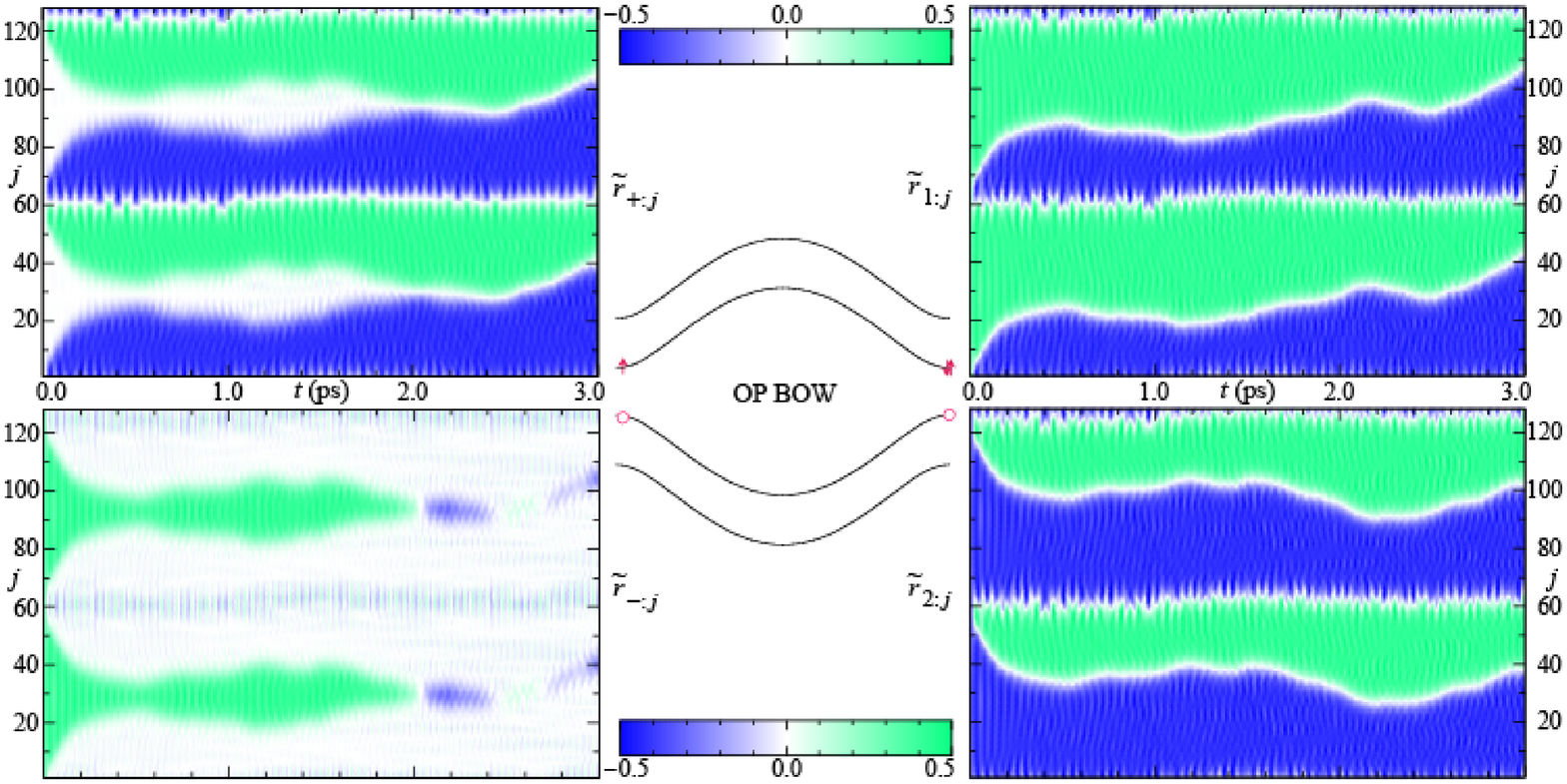}
\vspace*{-2mm}
\caption{(Color)
         The same as Fig. \ref{F:EPH4.8eVdynOP23-1} but in the case of a
         $2$-to-$3$ interband three-electron excitation at $k\simeq\pi$.}
\label{F:EPH4.8eVdynOP23-3}
\end{figure*}

\begin{figure*}
\centering
\includegraphics[width=170mm]{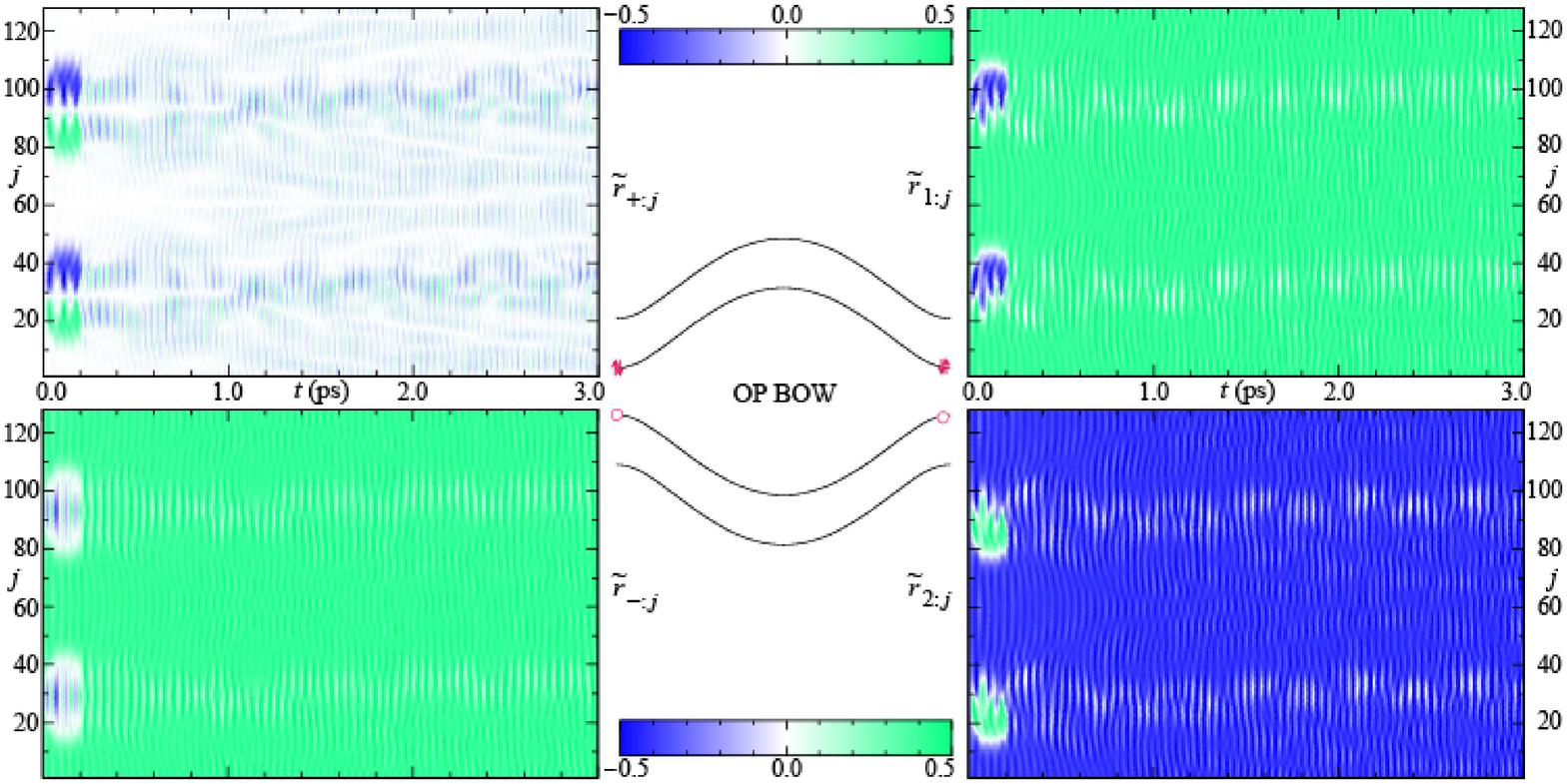}
\vspace*{-2mm}
\caption{(Color)
         The same as Fig. \ref{F:EPH4.8eVdynOP23-1} but in the case of a
         $2$-to-$3$ interband four-electron excitation at $k\simeq\pi$.}
\label{F:EPH4.8eVdynOP23-4}
\end{figure*}
\begin{figure*}
\centering
\includegraphics[width=170mm]{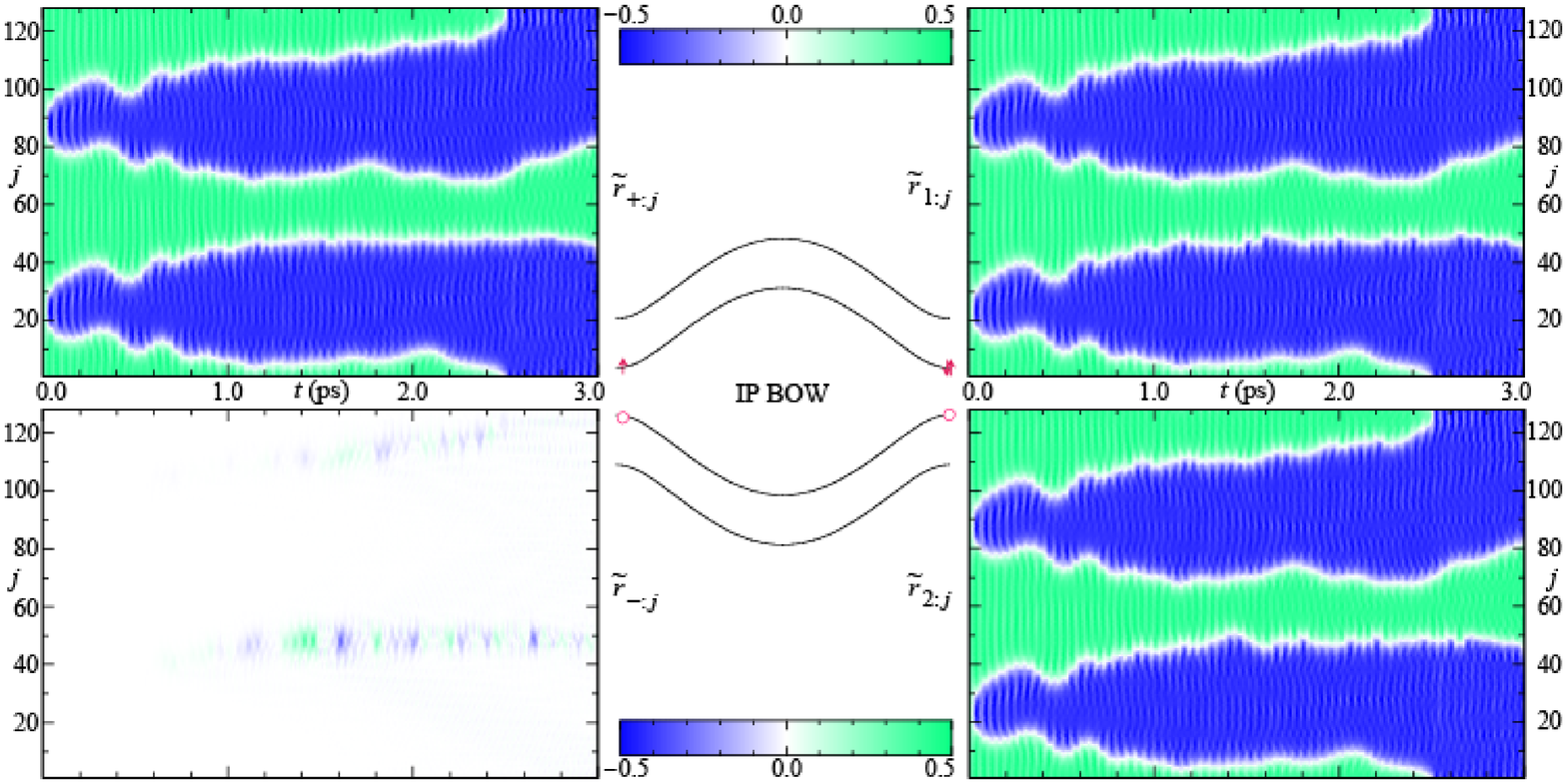}
\vspace*{-2mm}
\caption{(Color)
         The same as Fig. \ref{F:EPH4.8eVdynOP23-3} but on the IP-BOW
         background.}
\label{F:EPH4.8eVdynIP23-3}
\end{figure*}

\subsection{Correlated Electron-Lattice Dynamics}

   Now we switch on Coulomb interactions and discuss
correlated-electron-driven lattice dynamics in comparison with the
above.
When we excite one electron across the Peierls gap on the OP-BOW
background in the same manner that we did in Fig. \ref{F:SSHdynOP23-1},
the ground state remains silent without any explicit domain-wall
nucleation (Fig. \ref{F:EPH4.8eVdynOP23-1}).
With increasing number of absorbed photons, OP BOW happens to change
into IP BOW with three electrons being excited
(Fig. \ref{F:EPH4.8eVdynOP23-3})
but fails to do so even with four electrons being excited
(Fig. \ref{F:EPH4.8eVdynOP23-4}).
OP BOW seems to be necessarily converted into IP BOW with more than
four electrons being excited.
When electrons are correlated, an OP-to-IP conversion is not feasible
until the light irradiation comes up to a certain threshold intensity.
Even when an excitation density is given, the correlated dynamics still
depends crucially on how electrons are excited.
The initial excitation density is determined by the light intensity and
a threshold in it, if any, should originate from an energy barrier in
the relaxation path.
The critical number of excited electrons is not so decisive as it is
in a single chain, \cite{Y075113} because the spatial degrees of
freedom in charge-transfer excitations due to the ladder-like lattice
structure of polyacene may bypass the relaxation.
That is why the threshold excitation density is significantly
influenced by the microscopic information about photocarriers such as
their momenta and spins.

   What about photoexciting IP BOW under the influence of Coulomb
interactions?
Figure \ref{F:EPH4.8eVdynIP23-3} shows the case of three electrons
being excited.
The observations remain qualitatively unchanged from the uncorrelated
case.
It indeed happens that local OP-BOW domains appear with further
electrons being excited.
However, they hardly aggregate to develop a long-ranged order and
decline within several picoseconds.
Though the electronic correlation may stimulate local OP-BOW domains to
appear, yet a global IP-to-OP-BOW phase transition still looks hard to
photoinduce.
\begin{figure}
\centering
\includegraphics[width=80mm]{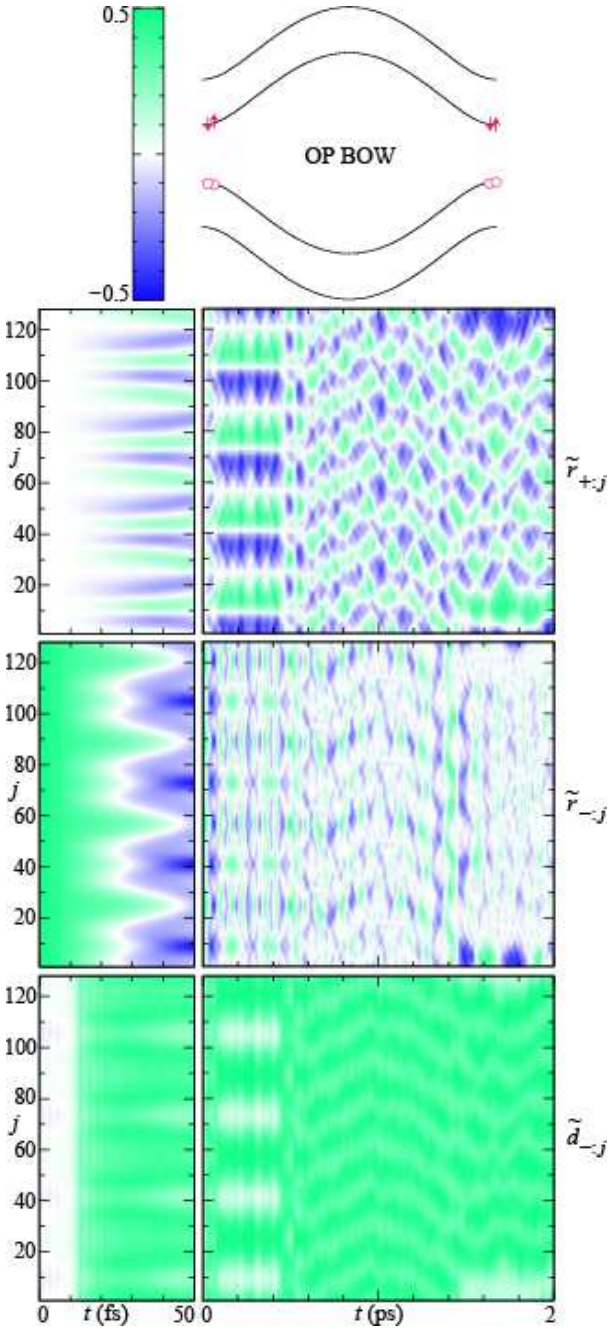}
\vspace*{-2mm}
\caption{(Color)
         $U=7.0\,\mbox{eV}$-EPH-model calculations of the bond
         variables $\widetilde{r}_{\pm:j}$ and the site variable
         $\widetilde{d}_{-:j}$ as functions of space $j$ and time $t$
         in the case of a $2$-to-$3$ interband four-electron excitation
         at $k\simeq\pi$ on the OP-BOW background.
         The initial $50$ femtoseconds are magnified in an attempt to
         visualize the ultrafast dynamics in more detail.}
\label{F:EPH7.0eVdynOP23-4}
\end{figure}

\begin{figure}
\centering
\includegraphics[width=80mm]{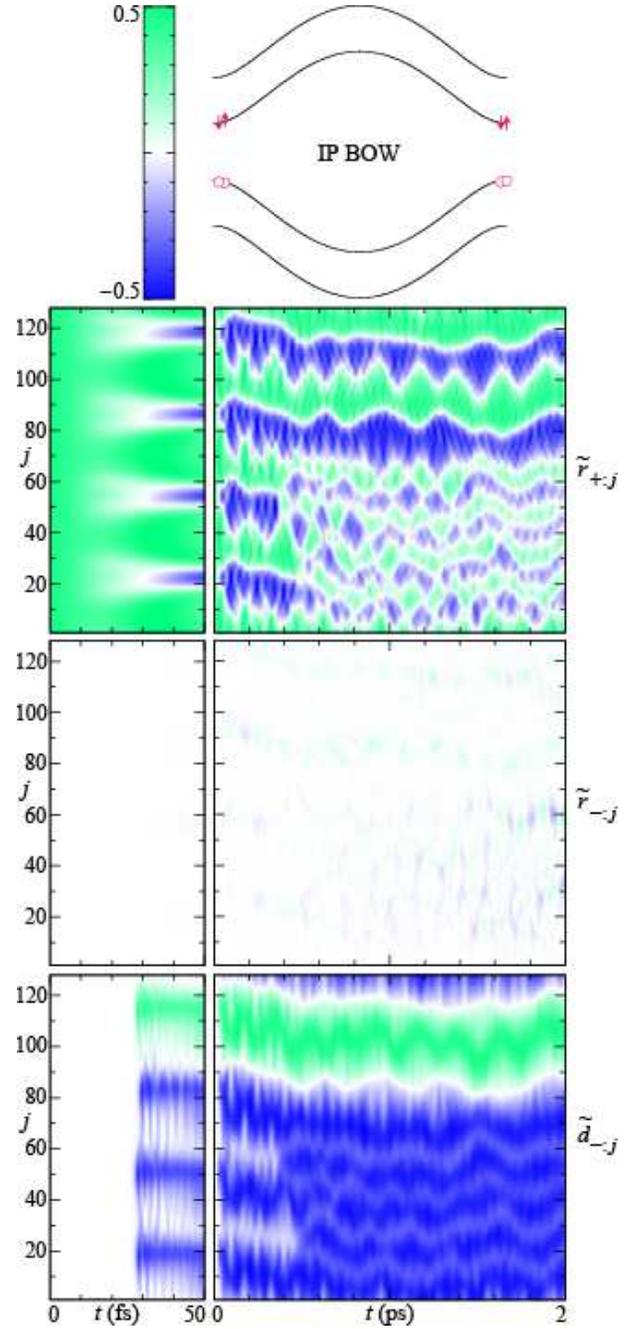}
\vspace*{-2mm}
\caption{(Color)
         The same as Fig. \ref{F:EPH7.0eVdynOP23-4} but on the IP-BOW
         background.}
\label{F:EPH7.0eVdynIP23-4}
\end{figure}

   Finally we increase the Coulomb interactions so as to get closer to
a phase boundary to CDW and photoexcite IP and OP BOWs by around
Peierls-gap energy.
Figure \ref{F:EPH7.0eVdynOP23-4} (\ref{F:EPH7.0eVdynIP23-4})
demonstrates the electron-lattice dynamics with four electrons being
excited from the OP-BOW (IP-BOW) ground state, where we define an
effective site variable,
$\widetilde{d}_{1:j}-\widetilde{d}_{2:j}\equiv\widetilde{d}_{-:j}$,
with $\widetilde{d}_{l:j}=(2d_{l:j}-d_{l:j-1}-d_{l:j+1})/4$ and plot it
as well as $\widetilde{r}_{\pm:j}$ in order to detect the CDW
configuration.
In both cases, electrons ultrafast form up into the CDW order and
then the BOW lattice arrangement melts down.
The {\it liquidized} BOW state should fade out if we include the effect
of extrinsic dissipation into our calculation.
A global and surviving CDW order is not attained until the
photoexcitation density reaches a certain threshold.
The development of the CDW order simply results from closing in upon
the phase boundary rather than increasing excitation energy, because
such observations are hardly obtained from every other previous
parametrization.
We have indeed tried various photoexcitations, including interband
electronic transitions of the $2$-to-$4$ type,  in both SSH and
$U=4.8\,\mbox{eV}$ EPH models, but such a global CDW formation is of no
occurrence.
It is not the BOW-to-CDW transition itself but lattice lagging behind
electrons in their combined dynamics that interests us here.
In this context we may be reminded of a charge-transfer compound,
tetrathiafulvalene-$p$-chloranil, where the electronic valence
instability and the dynamic lattice instability exhibit different time
scales. \cite{I057403}
First an ionic domain is photogenerated, whose rising time is so fast
as to be less than the experimental resolution, and then the
constituent molecules are stimulated to oscillate in a spin-Peierls-like
manner.
Both observations can be characterized as charge transfers preceding
lattice oscillations.
There is particularity in addition in the present findings.
Even in such ultrafast dynamics IP BOW gives an indication of its
superiority over OP BOW.
OP BOW is once rearranged into IP BOW and then liquidized, while
IP BOW directly breaks down without explicit nucleation of OP BOW.

\section{Summary and Discussion}

   We have optically characterized and manipulated the two isoenergetic
structures of polyacene with bond arrangements of the IP and OP types.
An OP-to-IP-BOW phase transition can be induced by the weakest
light irradiation unless electrons are correlated.
Coulomb interactions function as energy barriers in the relaxation path
and then the transition is conditioned on a good number of electrons
being excited.
On the other hand, an IP-to-OP-BOW phase transition is much less caused
by photoexcitations, whether electrons are correlated or not.
IP and OP BOWs may coexist in polyacene.
If the Peierls-distorted polyacene is irradiated with light of around
Peierls-gap energy, the mixed ground state is possibly unified into a
monostructure of the IP-BOW type.
The photoinduced predominant state can be annealed back into the
bimodal configuration.
Figure \ref{F:OC} shows that IP and OP BOWs selectively absorb
ultraviolet and blue-green rays, respectively, provided
$t_{\parallel}=2.4\,\mbox{eV}$.
Polyacene thus reveals itself as a photochromic polymer \cite{H1672}
with novel structural bistability.
\begin{figure*}
\centering
\includegraphics[width=170mm]{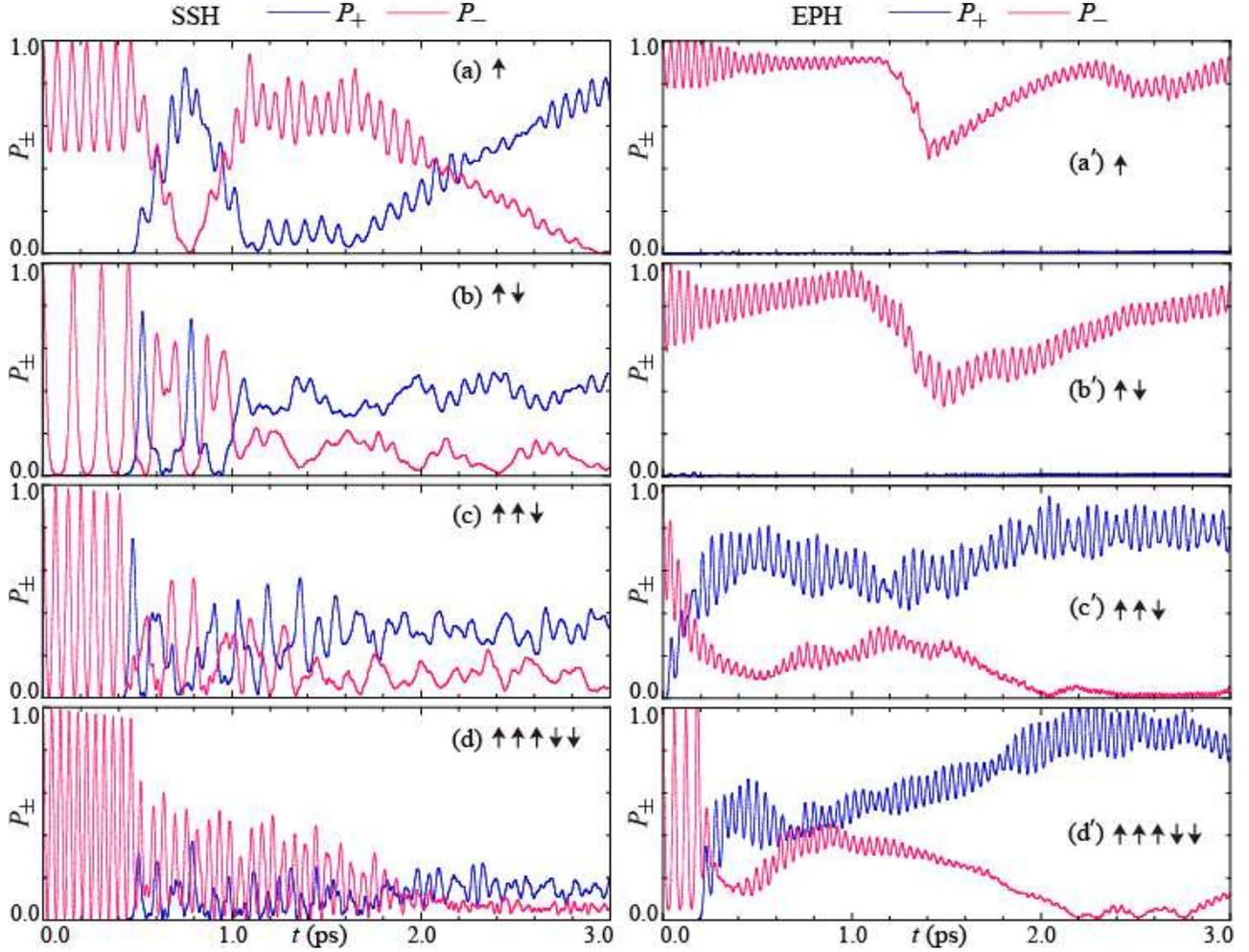}
\vspace*{-2mm}
\caption{(Color)
         Probabilities of finding the IP-BOW and OP-BOW configurations,
         $P_\pm$, as functions of time $t$ after photoexcitation:
         SSH-model [(a) to (d)] and
         $U=4.8\,\mbox{eV}$-EPH-model [(a$'$) to (d$'$)]
         calculations in the cases of $2$-to-$3$ interband
         one-electron [(a) and (a$'$)],
         two-electron [(b) and (b$'$)],
         three-electron [(c) and (c$'$)] and
         five-electron [(d) and (d$'$)] excitations
         at $k\simeq\pi$ on the OP-BOW background.
         (a) and (c) correspond to Figs. \ref{F:SSHdynOP23-1} and
         \ref{F:SSHdynOP23-3}, respectively, while
         (a$'$) and (c$'$) to Figs. \ref{F:EPH4.8eVdynOP23-1} and
         \ref{F:EPH4.8eVdynOP23-3}, respectively.}
\label{F:P+-}
\end{figure*}

   Photoinduced charge-transfer excitations potentially develop into
kink-antikink bound pairs.
Their breathing motions can cause a global phase change.
Figure \ref{F:EPH4.8eVdynOP23-4}, demonstrating geminate recombinations
of domain walls in the early stage, implies that correlated
electrons encounter energy barriers on the way of photogenerated
excitons being self-trapped and further dissociated into distant kinks
and antikinks.
A correlation-induced energy barrier on the relaxation path to far
distant domain walls is observed in platinum-halide chain compounds as
well, \cite{O045122,Y415215,M5758,I1088} which are well known to be
photoactive Peierls insulators.
Since a valley precedes the barrier, Coulomb interactions stimulate
the Frank-Condon state to relax and accelerate its localization.
\cite{Y415215}
We are thus convinced of the ultrafast quasifemtosecond
self-localization of photogenerated charge-transfer excitons in the
correlated dynamics
(Figs. \ref{F:EPH4.8eVdynOP23-3}--\ref{F:EPH7.0eVdynIP23-4}) and its
slowing down in the uncorrelated dynamics
(Figs. \ref{F:SSHdynOP23-1}--\ref{F:SSHdynIP24-2}).

   Coulomb interactions further serve quick and certain switching.
Comparing Figs. \ref{F:SSHdynOP23-3} and \ref{F:EPH4.8eVdynOP23-3},
we learn that the electronic correlation suppresses the macroscopic
oscillation of going back and forth between IP and OP BOWs.
In order to illuminate such a contrast between correlated and
uncorrelated electrons in more detail, we try to quantitatively
evaluate the weights of IP-BOW and OP-BOW configurations.
Their time-dependent and spatially-varying amplitudes may both be
normalized in comparison with the initial uniform BOW, because
coexistent stable IP-BOW and OP-BOW solutions have exactly the same
amplitude.
In the case of photoexciting OP BOW, the normalized momentary strengths
of IP-BOW and OP-BOW configurations can be defined as
$P_\pm(t)\equiv
 \sum_{j=1}^N \widetilde{r}_{\pm:j}^2(t)/
 \sum_{j=1}^N \widetilde{r}_{-:j}^2(0)$.
We calculate $P_\pm(t)$ with and without Coulomb interactions and plot
them in Fig. \ref{F:P+-}.
Besides the precursory global oscillation without any OP-to-IP-BOW
conversion yet, the initial OP-BOW state composed of uncorrelated
electrons repeats rise and fall on a macroscopic scale for a picosecond
or less.
The uncorrelated OP-BOW state is so persistent in its configuration.
On the other hand, the initial OP-BOW state of correlated electrons is
much more quickly and smoothly replaced by the photoinduced IP-BOW
phase once the excitation density reaches the threshold.
The promotion of OP-to-IP-BOW conversion is also recognizable as
a correlation effect.
With a few electrons being excited, the converted fraction amounts to
four fifths in the correlated case but stays less than two fifths in
the uncorrelated case.
Correlated electrons possess an adiabatic potential energy surface full
of ups and downs.
Once they climb over the energy barriers and find a stable
configuration, they are unlikely to swing back to the initial state.

   The macroscopic oscillation of uncorrelated electrons in their
photoconversion to the IP-BOW state and the correlation-driven
globalization of the IP-BOW domain are both understandable by
observing kink excitations in more detail.
The phase transition originates in generation of kink and antikink in
pairs.
Within the SSH modeling, neither spin nor charge accompanies any kink
as a result of resonating nonlinear excitations, \cite{Y415215,H2131}
where neutral and charged kinks are isoenergetic.
With increasing Coulomb interactions, their degeneracy is lifted and
those with excess spin (charge) density in their centers are relatively
stabilized (destabilized).
In the EPH calculations some of the kink excitations indeed bear net
spins and it is their intrachain interactions that develop and hold
the induced IP-BOW arrangement against the OP-BOW background.
In a successful OP-to-IP transition
[cf. Figs. \ref{F:EPH4.8eVdynOP23-3}, \ref{F:P+-}(c$'$), and
\ref{F:P+-}(d$'$)], we find two or more kinks conveying net spin
densities of the same sign in each chain, whereas in a frustrated one
[cf. Figs. \ref{F:EPH4.8eVdynOP23-1}, \ref{F:P+-}(a$'$), and
\ref{F:P+-}(b$'$)], we observe bound pairs of kink and antikink with
excess charge, rather than spin, densities disappear in the early stage.
The repulsive same-spin-carrying kinks may be the key to a global phase
change.
On the other hand, SSH kinks without any excess density of spin and
charge look hardly correlated and can therefore get closer to each
other (cf. Fig. \ref{F:SSHdynOP23-1}).
They are more and more created with increasing number of photons being
absorbed, but their geminate recombination is of frequent occurrence.
That is why the converted fraction $P_+$ oscillates on a macroscopic
scale in the SSH modeling [cf. Figs. \ref{F:P+-}(a)--\ref{F:P+-}(c) in
particular].
The reduction of both IP-BOW and OP-BOW strengths with increasing
excitation density [Figs. \ref{F:P+-}(a)--\ref{F:P+-}(d)] is also
attributable to the free kinks and thus-unlimited domain-wall
nucleation.

   A variety of photoinduced phase transitions have so far been
observed and analyzed indeed.
However, most of them assume the initial condition to be close to the
phase boundary and such a situation often depends in quite a critical
manner on some structural parameters which are hardly tractable.
Oligoacenes possess inherent bistability of topological origin and
their Peierls-distorted structural isomers are strictly and stably
isoenergetic.
Even under uniaxial pressure, for instance, applied in an attempt to
tune their intrinsic properties such as $t_\perp/t_\parallel$ and
$V_\perp/V_\parallel$, the strict bistability holds and enables us
to systematically investigate whether and how energetically degenerate
isomers are photoconverted into each other.
Polyacene derivatives and analogues such as polyphenanthrene
\cite{T1069} and polyacenacene \cite{Y823} will make possible further
comparative studies.
We are looking forward to more and more interest in this potential
organic polymer.

\acknowledgments

   The author is grateful to J. Ohara for useful comments.
This work was supported by the Suhara Memorial Foundation and
the Ministry of Education, Culture, Sports,
Science, and Technology of Japan.

\end{document}